\documentclass[aps, prl,twocolumn, 10pt, showpacs, longbibliography, superscriptaddress]{revtex4-1}
\usepackage{graphicx}% Include figure files
\usepackage{dcolumn}% Align table columns on decimal point
\usepackage{bm}% bold math
\usepackage{color}
\usepackage{amsmath, amsthm, amssymb}
\usepackage[modulo]{lineno}%[switch]{lineno} 
\usepackage[usenames,dvipsnames]{xcolor}
\newcommand{\ba}{\begin{eqnarray}}
\newcommand{\ea}{\end{eqnarray}}

\usepackage[colorlinks=true, urlcolor=violet, linkcolor=blue, citecolor=red, hyperindex=true, linktocpage=true]{hyperref}

\usepackage{float}
\floatstyle{boxed}

\begin{document}
\title{Observation of the Dirac fluid\\ and the breakdown of the Wiedemann-Franz law in graphene}
%% Notice placement of commas and superscripts and use of &
%% in the author list
\author{Jesse Crossno}
\affiliation{Department of Physics, Harvard University, Cambridge, MA 02138, USA}
\affiliation{John A. Paulson School of Engineering and Applied Sciences, Harvard University, Cambridge, MA 02138, USA}
\author{Jing K. Shi}
\affiliation{Department of Physics, Harvard University, Cambridge, MA 02138, USA}
\author{Ke Wang}
\affiliation{Department of Physics, Harvard University, Cambridge, MA 02138, USA}
\author{Xiaomeng Liu}
\affiliation{Department of Physics, Harvard University, Cambridge, MA 02138, USA}
\author{Achim Harzheim}
\affiliation{Department of Physics, Harvard University, Cambridge, MA 02138, USA}
\author{Andrew Lucas}
\affiliation{Department of Physics, Harvard University, Cambridge, MA 02138, USA}
\author{Subir Sachdev}
\affiliation{Department of Physics, Harvard University, Cambridge, MA 02138, USA}
\affiliation{Perimeter Institute for Theoretical Physics, Waterloo, Ontario N2L 2Y5, Canada}
\author{Philip Kim}
\email{pkim@physics.harvard.edu}
\affiliation{Department of Physics, Harvard University, Cambridge, MA 02138, USA}
\affiliation{John A. Paulson School of Engineering and Applied Sciences, Harvard University, Cambridge, MA 02138, USA}
\author{Takashi Taniguchi}
\affiliation{National Institute for Materials Science, Namiki 1-1, Tsukuba, Ibaraki 305-0044, Japan}
\author{Kenji Watanabe}
\affiliation{National Institute for Materials Science, Namiki 1-1, Tsukuba, Ibaraki 305-0044, Japan}
\author{Thomas A. Ohki}
\affiliation{Raytheon BBN Technologies, Quantum Information Processing Group, Cambridge,
Massachusetts 02138, USA}
\author{Kin Chung Fong}
\email{kc.fong@bbn.com}
\affiliation{Raytheon BBN Technologies, Quantum Information Processing Group, Cambridge,
Massachusetts 02138, USA}
\date{\today}
\begin{abstract}%
Interactions between particles in quantum many-body systems can lead to collective behavior described by hydrodynamics. One such system is the electron-hole plasma in graphene near the charge neutrality point which can form a strongly coupled Dirac fluid. This charge neutral plasma of quasi-relativistic fermions is expected to exhibit a substantial enhancement of the thermal conductivity, due to decoupling of charge and heat currents within hydrodynamics.   Employing high sensitivity Johnson noise thermometry, we report the breakdown of the Wiedemann-Franz law in graphene, with a thermal conductivity an order of magnitude larger than the value predicted by Fermi liquid theory.  This result is a signature of the Dirac fluid, and constitutes direct evidence of collective motion in a quantum electronic fluid.
\end{abstract}
%\pacs{65.80.Ck, 68.65.-k, and 07.57.Kp}
\maketitle 
%\linenumbers
%\setstretch{2}
%\doublespacing
Understanding the dynamics of many interacting particles is a formidable task in physics, complicated by many coupled degrees of freedom.   For electronic transport in matter, strong interactions can lead to a breakdown of the Fermi liquid (FL) paradigm of coherent quasiparticles scattering off of impurities.  In such situations, the complex microscopic dynamics can be coarse-grained to a hydrodynamic description of momentum, energy, and charge transport on long length and time scales \cite{Kadanoff:1963jv}. Hydrodynamics has been successfully applied to a diverse array of interacting quantum systems, from high mobility electrons in conductors \cite{DeJong:1995bn}, to cold atoms \cite{Cao07012011} and quark-gluon plasmas \cite{shuryak}.   As has been argued for strongly interacting massless Dirac fermions in graphene at the charge-neutrality point (CNP) \cite{Muller:2008ud, Foster:2009vl, Apostolov:2014vo, Narozhny:2015vc}, hydrodynamic effects are expected to greatly modify transport coefficients as compared to their FL counterparts.

Many-body physics in graphene is interesting due to electron-hole symmetry and a linear dispersion relation at the CNP \cite{Novoselov:2005ix, Zhang:2005gp}. In particular, the Fermi surface vanishes, leading to ineffective screening \cite{Siegel:2013wh} and the formation of a strongly-interacting quasi-relativistic electron-hole plasma, known as a Dirac fluid \cite{Sheehy:2007bi}.  The Dirac fluid shares many features with quantum critical systems \cite{Keimer:2011ug}: most importantly, the electron-electron scattering time is fast \cite{Lui:2010kh, Breusing:2009ba, koppensNP, johannsen}, and well suited to a hydrodynamic description.   A number of exotic properties have been predicted including nearly perfect (inviscid) flow \cite{Muller:2009cy} and a diverging thermal conductivity resulting in the breakdown of the Wiedemann-Franz law \cite{Muller:2008ud, Foster:2009vl}.

\begin{figure*}
\includegraphics[width=6.25in]{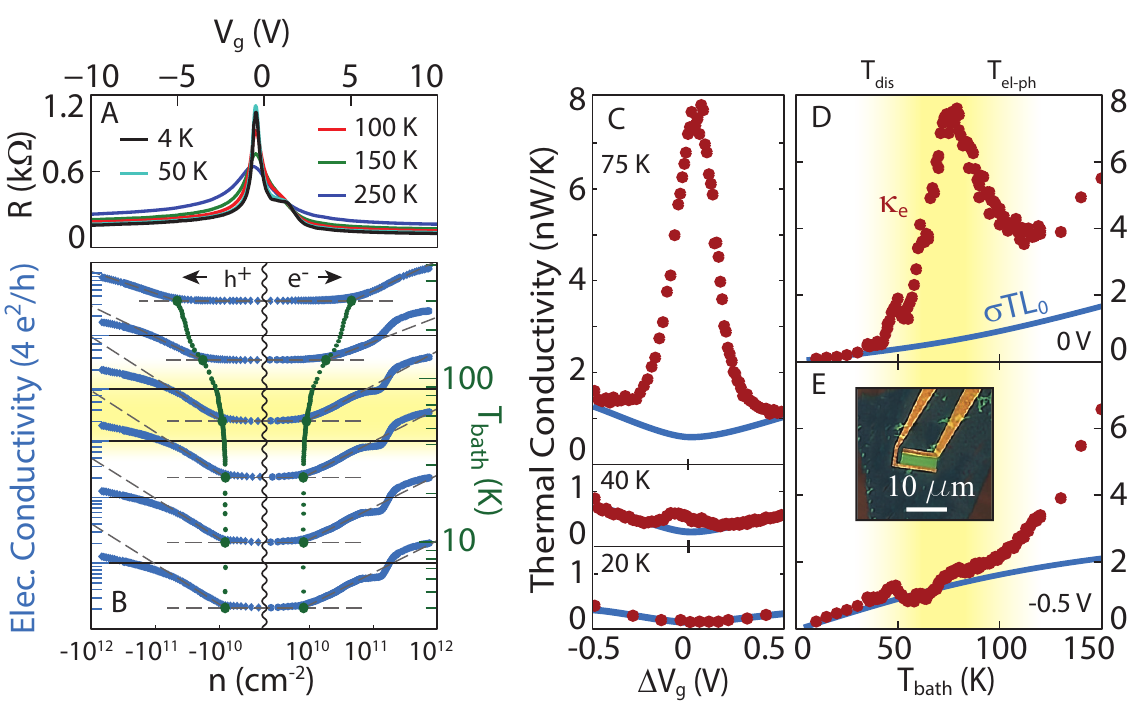}
\caption{\textbf{Temperature and density dependent electrical and thermal conductivity.} \textbf{(A)} Resistance versus gate voltage at various temperatures. \textbf{(B)} Electrical conductivity (blue) as a function of the charge density set by the back gate for different bath temperatures. The residual carrier density at the neutrality point (green) is estimated by the intersection of the minimum conductivity with a linear fit to $\log(\sigma)$ away from neutrality (dashed grey lines). Curves have been offset vertically such that the minimum density (green) aligns with the temperature axis to the right. Solid black lines correspond to $4e^2/h$. At low temperature, the minimum density is limited by disorder (charge puddles). However, above $T_{\mathrm{dis}}\sim$ 40~K, a crossover marked in the half-tone background, thermal excitations begin to dominate and the sample enters the non-degenerate regime near the neutrality point. \textbf{(C-D)} Thermal conductivity (red points) as a function of (C) gate voltage and (D) bath temperature compared to the Wiedemann-Franz law, $\sigma T\mathcal{L}_0$ (blue lines).  At low temperature and/or high doping ($|\mu| \gg k_{\mathrm{B}}T$), we find the WF law to hold.  This is a non-trivial check on the quality of our measurement. In the non-degenerate regime ($|\mu|<k_{\mathrm{B}}T$) the thermal conductivity is enhanced and the WF law is violated. Above $T_{\mathrm{el-ph}}\sim$ 80~K, electron-phonon coupling becomes appreciable and begins to dominate thermal transport at all measured gate voltages.   All data from this figure is taken from sample S2 (inset 1E).}
\label{Fig1}
\end{figure*}

Away from the CNP, graphene has a sharp Fermi surface and the standard Fermi liquid (FL) phenomenology holds. By tuning the chemical potential, we may measure thermal and electrical conductivity in both the Dirac fluid (DF) and the FL in the same sample. In a FL, the relaxation of heat and charge currents is closely related as they are carried by the same quasiparticles. The Wiedemann-Franz (WF) law \cite{Ashcroft:ud} states that the electronic contribution to a metal's thermal conductivity $\kappa_{\mathrm{e}}$ is proportional to its electrical conductivity $\sigma$ and temperature $T$, such that the Lorenz ratio $\mathcal{L}$ satisfies
\begin{equation}
\label{WF}
\mathcal{L}\equiv\frac{\kappa_{\mathrm{e}}}{\sigma T}=\frac{\pi^2}{3}\left(\frac{k_{\mathrm{B}}}{e}\right)^2\equiv\mathcal{L}_0
\end{equation}
where $e$ is the electron charge, $k_{\mathrm{B}}$ is the Boltzmann constant, and $\mathcal{L}_0$ is the Sommerfeld value derived from FL theory. $\mathcal{L}_0$ depends only on fundamental constants, and not on specific details of the system such as carrier density or effective mass. As a robust prediction of FL theory, the WF law has been verified in numerous metals \cite{Ashcroft:ud}. However, in recent years, an increasing number of non-trivial violations of the WF law have been reported in strongly interacting systems such as Luttinger liquids \cite{Wakeham:2011dp}, metallic ferromagnets \cite{Smith:2008kv}, heavy fermion metals \cite{Pfau:2012hc}, and underdoped cuprates \cite{Hill:2001gf}, all related to the emergence of non-Fermi liquid behavior.

The WF law is expected to be violated at the CNP in a DF due to the strong Coulomb interactions between thermally excited charge carriers. An electric field drives electrons and holes in opposite directions; collisions between them introduce a frictional dissipation, resulting in a finite conductivity even in the absence of disorder \cite{Fritz:2008go}. In contrast, a temperature gradient causes electrons and holes to move in the same direction inducing an energy current, which grows unimpeded by inter-particle collisions (Fig. 3C inset). The thermal conductivity is therefore limited only by the rate at which momentum is relaxed due to residual impurities.

Realization of the Dirac fluid in graphene requires that the thermal energy be larger than the local chemical potential $\mu(\mathbf{r})$, defined at position $\mathbf{r}$: $k_{\mathrm{B}}T\gtrsim |\mu(\mathbf{r})|$. Impurities cause spatial variations in the local chemical potential, and even when the sample is globally neutral, it is locally doped to form electron-hole puddles with finite $\mu(\mathbf{r})$ \cite{Adam:2007kc, Martin:2008uf, Zhang:2009ce, Xue:2011kg}. Formation of the DF is further complicated by phonon scattering at high temperature which can relax momentum by creating additional inelastic scattering channels. This high temperature limit occurs when the electron-phonon scattering rate becomes comparable to the electron-electron scattering rate. These two temperatures set the experimental window in which the DF and the breakdown of the WF law can be observed.

To minimize disorder, the monolayer graphene samples used in this report are encapsulated in hexagonal boron nitride (hBN) \cite{Dean:2010jy}. All devices used in this study are two-terminal to keep a well-defined temperature profile \cite{Fong:2012ut} with contacts fabricated using the one-dimensional edge technique \cite{Wang:2013tl} in order to minimize contact resistance. We employ a back gate voltage $V_{\mathrm{g}}$ applied to the silicon substrate to tune the charge carrier density $n=n_{\mathrm{e}}-n_{\mathrm{h}}$, where $n_{\mathrm{e}}$ and $n_{\mathrm{h}}$ are the electron and hole density, respectively (see supplementary materials (SM)). All measurements are performed in a cryostat controlling the temperature $T_{\mathrm{bath}}$. Fig. 1A shows the resistance $R$ versus $V_{\mathrm{g}}$ measured at various fixed temperatures for a representative device (see SM for all samples). From this, we estimate the electrical conductivity $\sigma$ (Fig.~1B) using the known sample dimensions. At the CNP, the residual charge carrier density $n_{\mathrm{min}}$ can be estimated by extrapolating a linear fit of $\log(\sigma)$ as a function of $\log(n)$ out to the minimum conductivity \cite{Couto:2014ip}. At the lowest temperatures we find $n_{\mathrm{min}}$ saturates to $\sim$8$\times$10$^9~$cm$^{-2}$. We note that the extraction of $n_{min}$ by this method overestimates the charge puddle energy, consistent with previous reports \cite{Dean:2010jy}. Above the disorder energy scale $T_{\mathrm{dis}}\sim$40~K, $n_{\mathrm{min}}$ increases as $T_{\mathrm{bath}}$ is raised, suggesting thermal excitations begin to dominate and the sample enters the non-degenerate regime near the CNP.

The electronic thermal conductivity is measured using high sensitivity Johnson noise thermometry (JNT) \cite{Fong:2012ut, Crossno:2015ez}. We apply a small bias current through the sample that injects a joule heating power $P$ directly into the electronic system, inducing a small temperature difference $\Delta T\equiv T_{\mathrm{e}}-T_{\mathrm{bath}}$ between the graphene electrons and the bath. The electron temperature $T_{\mathrm{e}}$ is monitored independent of the lattice temperature through the Johnson noise power emitted at 100~MHz with a 20~MHz bandwidth defined by an LC matching network. We designed our JNT to be operated over a wide temperature range 3--300~K \cite{Crossno:2015ez}. With a precision of $\sim 10$ mK, we measure small deviations of $T_{\mathrm{e}}$ from $T_{\mathrm{bath}}$, i.e. $\Delta T\ll T_{\mathrm{bath}}$. In this limit, the temperature of the graphene lattice is well thermalized to the bath \cite{Fong:2012ut} and our JNT setup allows us to sensitively measure the electronic cooling pathways in graphene. At low enough temperatures, electron and lattice interactions are weak \cite{Crossno:2015ez, Fong:2013hl}, and most of the Joule heat generated in graphene escapes via direct diffusion to the contacts (SM). As temperature increases, electron-phonon scattering becomes appreciable and thermal transport becomes limited by the electron-phonon coupling strength \cite{Fong:2013hl, Betz:2012wya, McKitterick:2015vc}. The onset temperature of appreciable electron-phonon scattering, $T_{\mathrm{el-ph}}$, depends on the sample disorder and device geometry: $T_{\mathrm{el-ph}}\sim$80~K \cite{Fong:2013hl, Crossno:2015ez, Yigen:2014vw, hakonen} for our samples. Below this temperature, the electronic contribution of the thermal conductivity can be obtained from $P$ and $\Delta T$ using the device dimensions (SM).

Fig. 1C plots $\kappa_{\mathrm{e}}(V_{\mathrm{g}})$ alongside the simultaneously measured $\sigma(V_{\mathrm{g}})$ at various fixed bath temperatures. Here, for a direct quantitative comparison based on the WF law, we plot the scaled electrical conductivity as $\sigma T\mathcal{L}_0$ in the same units as $\kappa_{\mathrm{e}}$. At low temperatures, $T<T_{\mathrm{dis}}\sim$ 40~K, where the puddle induced density fluctuations dominates, we find $\kappa_{\mathrm{e}}\approx\sigma T\mathcal{L}_0$, monotonically increasing as a function of carrier density with a minimum at the neutrality point, confirming the WF law in the disordered regime. As $T$ increases ($T>T_{\mathrm{dis}}$), however, the measured $\kappa_{\mathrm{e}}$ begins to deviate from the FL theory. We note that this violation of the WF law only appears close to the CNP, with the measured thermal conductivity maximized at $n=0$ (Fig 1D). % in the temperature range $T{\mathrm{dis}}<T<T_{\mathrm{e-ph}}$. 
The deviation is the largest at  75~K, where $\kappa_{\mathrm{e}}$ is over an order of magnitude larger than the value expected for a FL.   This non-FL behavior quickly disappears as $|n|$ increases; $\kappa_{\mathrm{e}}$ returns to the FL value and restores the WF law.   In fact, away from the CNP, the WF law holds for a wide temperature range, consistent with previous reports \cite{Crossno:2015ez, Yigen:2014vw, Fong:2013hl} (Fig. 1E).    For this FL regime, we verify the WF law up to $T_{\mathrm{el-ph}} \sim 80$ K.  Finally, in the high temperature regime $T>T_{\mathrm{el-ph}}$, the additional electron-phonon cooling pathway causes the measured thermal conductivity to be larger than $\kappa_{\mathrm{e}}$.  We find that near the CNP $\kappa_{\mathrm{e}}$ tends to decrease just before $T_{\mathrm{el-ph}}$, restricting the maximal observable violation of the WF law.

\begin{figure}
\includegraphics[width=\columnwidth]{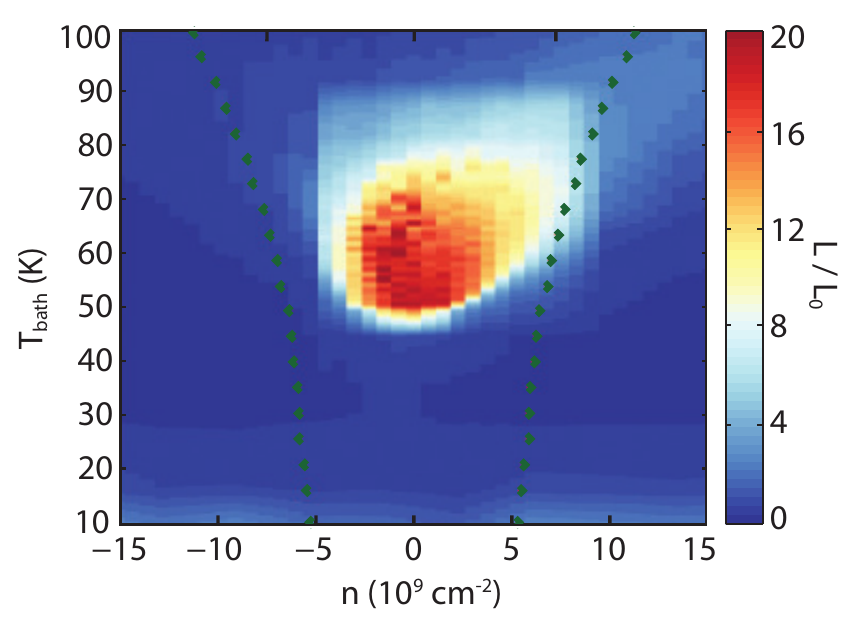}
\caption{\textbf{Breakdown of the Wiedemann-Franz law in the Dirac fluid regime.} The Lorenz ratio is shown as a function of the charge carrier density and bath temperature. Near the CNP and for temperatures above the disorder (charge puddle) regime but below the onset of electron-phonon coupling, the Lorenz ratio is measured to be an order of magnitude greater than the Fermi liquid value of 1 (blue). The WF law is observed to hold outside of the Dirac fluid regime.  All data from this figure is taken from sample S1.}
\label{Fig2}
\end{figure}

Our observation of the breakdown of the WF law in graphene is consistent with the emergence of the DF. Fig 2 shows the full density and temperature dependence of the experimentally measured Lorenz ratio in order to highlight the presence of the DF. The blue colored region denotes $\mathcal{L}\sim\mathcal{L}_0$, suggesting the carriers in graphene exhibit FL behavior. The WF law is violated in the DF (yellow-red) with a peak Lorenz ratio 22 times larger than $\mathcal{L}_0$. The green dotted line shows the corresponding $n_{\mathrm{min}}(T)$ for this sample; the DF is found within this regime, indicating the coexistence of thermally populated electrons and holes. We find that disorder and phonon scattering bound the temperature range of the Dirac fluid, $T_{\mathrm{dis}}<T<T_{\mathrm{el-ph}}$.

We investigate the effect of impurities on hydrodynamic transport by comparing the results obtained from samples with varying disorder. Fig. 3A shows $n_{\mathrm{min}}$ as a function of temperature for three samples used in this study. $n_{\mathrm{min}}(T=0)$ is estimated as 5, 8, and 10$\times$10$^9~$cm$^{-2}$ in samples S1, S2, and S3, respectively. All devices show qualitatively similar Dirac fluid behavior; the largest value of $\mathcal{L}/\mathcal{L}_0$ measured in the Dirac fluid regime is 22, 12 and 3 in samples S1, S2, and S3, respectively (Fig 3B). For a direct comparison, we show $\mathcal{L}(n)$ for all three samples at the same temperature (60~K) in Fig 3C. We find that cleaner samples not only have a more pronounced peak but also a narrower density dependence, as predicted \cite{Muller:2008ud, Foster:2009vl}.

\begin{figure}
\includegraphics[width=\columnwidth]{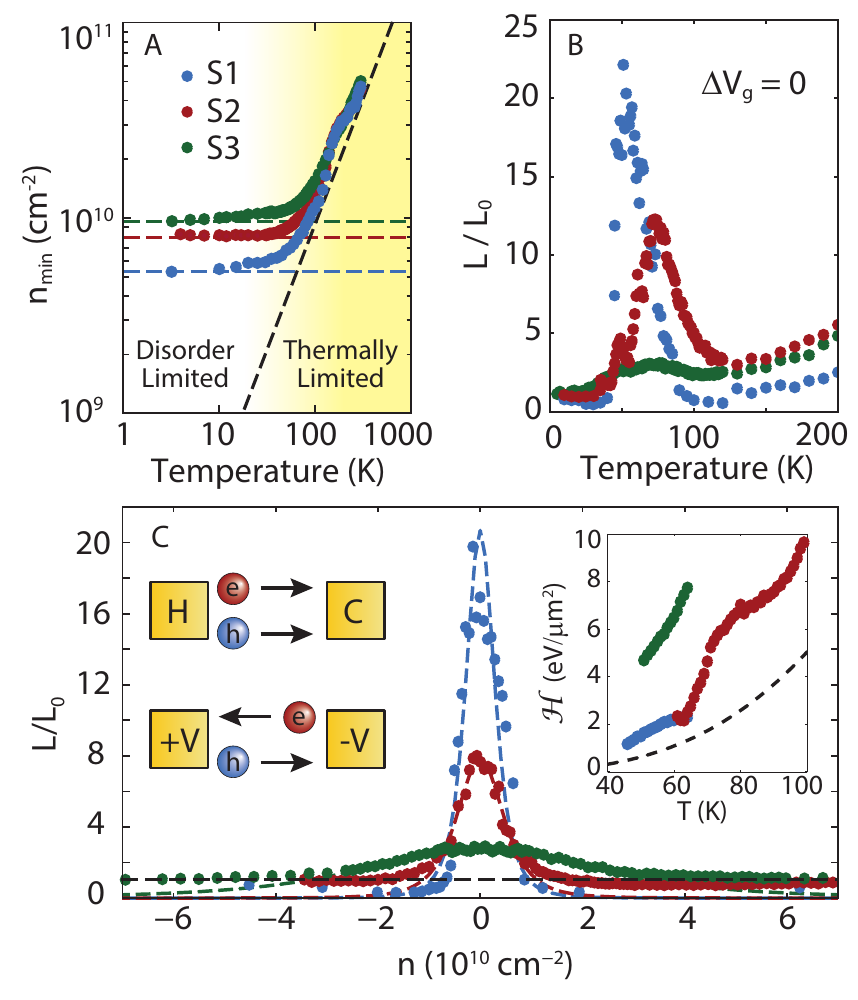}
\caption{\textbf{Disorder in the Dirac fluid.  (A)} Minimum carrier density as a function of temperature for all three samples.  At low temperature each sample is limited by disorder. At high temperature all samples become limited by thermal excitations.  Dashed lines are a guide to the eye. \textbf{(B)} The Lorentz ratio of all three samples as a function of bath temperature. The largest WF violation is seen in the cleanest sample. \textbf{(C)} The gate dependence of the Lorentz ratio is well fit to hydrodynamic theory of Ref. \cite{Muller:2008ud, Foster:2009vl}. Fits of all three samples are shown at 60~K. All samples return to the Fermi liquid value (black dashed line) at high density. Inset shows the fitted enthalpy density as a function of temperature and the theoretical value in clean graphene (black dashed line). Schematic inset illustrates the difference between heat and charge current in the neutral Dirac plasma.}
\label{Fig3}
\end{figure}

More quantitative analysis of $\mathcal{L}(n)$ in our experiment can be done by employing a quasi-relativistic hydrodynamic theory of the DF incorporating the effects of weak impurity scattering \cite{Hartnoll:2007ub, Muller:2008ud, Foster:2009vl}. 
\begin{equation}
\label{Lorentz}
 \mathcal{L} =  \frac{\mathcal{L}_{\mathrm{DF}}}{\left(1+(n/n_0)^2\right)^2}
 \end{equation}
 where 
\begin{equation}
\mathcal{L}_{\mathrm{DF}} = \frac{\mathcal{H}v_{\mathrm{F}}l_{\mathrm{m}}}{T^2 \sigma_{\mathrm{min}}} \;\;\;\; \textrm{~and~} \;\;\;\; n_0^2 = \frac{\mathcal{H}\sigma_{\mathrm{min}}}{e^2v_{\mathrm{F}}l_{\mathrm{m}}}.
\end{equation}
Here $v_{\mathrm{F}}$ is the Fermi velocity in graphene, $\sigma_{\mathrm{min}}$ is the electrical conductivity at the CNP, $\mathcal{H}$ is the fluid enthalpy density, and $l_m$ is the momentum relaxation length from impurities. Two parameters in Eqn. (\ref{Lorentz}) are undetermined for any given sample: $l_m$ and  $\mathcal{H}$.  For simplicity, we assume we are well within the DF limit where $l_m$ and $\mathcal{H}$ are approximately independent of $n$. We fit Eqn. (\ref{Lorentz}) to the experimentally measured $\mathcal{L}(n)$ for all temperatures and densities in the Dirac fluid regime to obtain $l_{\mathrm{m}}$ and $\mathcal{H}$  for each sample.  Fig 3C shows three representative fits to Eqn. (\ref{Lorentz}) taken at 60 K. $l_{\mathrm{m}}$ is estimated to be 1.5, 0.6, and 0.034 $\mu$m for samples S1, S2, and S3, respectively.  For the system to be well described by hydrodynamics, $l_{\mathrm{m}}$ should be long compared to the electron-electron scattering length of $\sim$0.1 $\mu$m expected for the Dirac fluid at 60 K \cite{Muller:2009cy}.  This is consistent with the pronounced signatures of hydrodynamics in S1 and S2, but not in S3, where only a glimpse of the DF appears in this more disordered sample. Our analysis also allows us to estimate the thermodynamic quantity $\mathcal{H}(T)$ for the DF. The Fig. 3C inset shows the fitted enthalpy density as a function of temperature compared to that expected in clean graphene (dashed line) \cite{Muller:2009cy}, excluding renormalization of the Fermi velocity. In the cleanest sample $\mathcal{H}$ varies from 1.1-2.3 eV/$\mu$m$^2$ for $T_{\mathrm{dis}}<T<T_{\mathrm{el-ph}}$.   This enthalpy density corresponds to  $\sim$ 20 meV or $\sim4k_{\mathrm{B}}T$ per charge carrier --- about a factor of 2 larger than the model calculation without disorder \cite{Muller:2009cy}.  

To fully incorporate the effects of disorder, a hydrodynamic theory treating inhomogeneity non-perturbatively  is necessary  \cite{lucas, lucas2}. The enthalpy densities reported here are larger than the theoretical estimation obtained for disorder free graphene, consistent with the picture that chemical potential fluctuations prevent the sample from reaching the Dirac point. 
While we find thermal conductivity well described by Ref.~\cite{Muller:2008ud, Foster:2009vl}, electrical conductivity increases slower than expected away from the CNP,  a result consistent with hydrodynamic transport in a viscous fluid with charge puddles \cite{lucas2}. 

In a hydrodynamic system, the ratio of shear viscosity $\eta$ to entropy density $s$ is an indicator of the strength of the interactions between constituent particles.   It is suggested that the DF can behave as a nearly perfect fluid \cite{Muller:2009cy}: $\eta/s$ approaches a conjecture by Kovtun-Son-Starinets: $(\eta/s)/(\hbar/k_B) \gtrsim 1/4\pi$ for a strongly interacting system \cite{Kovtun:2005tz}.  A non-perturbative hydrodynamic framework can be employed to estimate $\eta$, as we discuss elsewhere \cite{lucas2}.   A direct measurement of $\eta$ is of great interest.

%An extended theory and further experiments may explain this behavior and further refine our estimation of viscosity.

We have experimentally discovered the breakdown of the WF law and provided evidence for the hydrodynamic behavior of the Dirac fermions in graphene. This provides an experimentally realizable Dirac fluid and opens the way for future studies of strongly interacting relativistic many-body systems. Beyond a diverging thermal conductivity and an ultra-low viscosity, other peculiar phenomena are expected to arise in this plasma. The massless nature of the Dirac fermions is expected to result in a large kinematic viscosity, despite  a small shear viscosity $\eta$.   Observable hydrodynamic effects have also been predicted to extend into the FL regime \cite{principi}.  The study of magnetotransport in the DF will lead to further tests of hydrodynamics \cite{Muller:2008ud,Hartnoll:2007ub}.

\textbf{Acknowledgements.}   We thank Matthew Foster and Dmitri Efetov for helpful discussions. The major experimental work at Harvard is supported by DOE (DE-SC0012260) and at Raytheon BBN Technologies is supported by IRAD.  J.C. thanks the support of the FAME Center, sponsored by SRC MARCO and DARPA. K.W. is supported by ARO MURI (W911NF-14-1-0247). J.K.S. is supported by ARO (W911NF-14-1-0638) and AStar. P.K. acknowledges partial support from the Gordon and Betty Moore Foundation's EPiQS Initiative through Grant GBMF4543 and Nano Material Technology Development Program through the National Research Foundation of Korea (2012M3A7B4049966).  A.L. and S.S. are supported by the NSF under Grant DMR-1360789, the Templeton foundation, and MURI grant W911NF-14-1-0003 from ARO.
Research at Perimeter Institute is supported by the Government of Canada through Industry Canada 
and by the Province of Ontario through the Ministry of Research and Innovation.  K.W. and T.T. acknowledge support from the Elemental Strategy Initiative conducted by the MEXT, Japan. T.T. acknowledges support from a Grant-in-Aid for Scientific Research on Grant 262480621 and on Innovative Areas ``Nano Informatics" (Grant 25106006) from JSPS. T.A.O. and K.C.F. acknowledge Raytheon BBN Technologies' support for this work.  This work was performed in part at the Center for Nanoscale Systems (CNS), a member of the National Nanotechnology Infrastructure Network (NNIN), which is supported by the National Science Foundation under NSF award no. ECS-0335765. CNS is part of Harvard University.

%\pagebreak

\part*{Supplementary Materials}
\setcounter{totalnumber}{5}
\renewcommand\textfraction{.1}

\newsavebox{\foobox}
\newcommand{\slantbox}[2][0]{\mbox{%
        \sbox{\foobox}{#2}%
        \hskip\wd\foobox
        \pdfsave
        \pdfsetmatrix{1 0 #1 1}%
        \llap{\usebox{\foobox}}%
        \pdfrestore
}}
\newcommand\unslant[2][-.25]{\slantbox[#1]{$#2$}}

\newcommand{\mmu}{\text{\unslant\mu}}
\newcommand{\mpi}{\text{\unslant[-.18]\pi}}
\newcommand{\mdelta}{\text{\unslant[-.18]\delta}}
\newcommand{\mzeta}{\text{\unslant[-.15]\zeta}}
\makeatletter
\renewcommand{\theequation}{S.\arabic{equation}}
\renewcommand{\thefigure}{S\arabic{figure}} 
\renewcommand{\thetable}{S.\Roman{table}} 
\renewcommand{\thesection}{S-\Roman{section}} 
\makeatother

\setcounter{figure}{0}
\setcounter{equation}{0}

\section{Sample Fabrication}
Single layer graphene is encapsulated in hexagonal boron nitride on an n-doped silicon wafer with 285 nm SiO$_2$ \cite{Wang:2013tl} and is subsequently annealed in vacuum for 15 minutes at 350 $^\circ$C. It is then etched using reactive-ion-etching (RIE) to define the width of the device. A second etch mask is then lithographically defined to overlap with the sample edge, leaving the rest of the sample rectangular shaped with the desired aspect ratio. After the RIE is performed, the same etch mask is used as the metal deposition mask, upon which Cr/Pd/Au (1.5 nm / 5 nm / 200 nm) is deposited. The resulting Ohmic contacts show low contact resistances and small PN junction effects due to their minimum overlap with device edge. 

\section{Optimizing samples for high frequency thermal conductivity measurements}
To measure the electronic thermal conductivity $\kappa_{\mathrm{e}}$ of graphene using high frequency Johnson noise the sample design should be made with three additional considerations: stray chip capacitance, resistance of the lead wires, and sample dimensions that enhance electron diffusion cooling over phonon coupling. 

Johnson noise thermometry (JNT) relies on measuring the total noise power emitted in a specified frequency band and relating that to the electronic temperature on the device; to maximize the sensitivity, high frequency and wide bandwidth measurements should be made \cite{Crossno:2015ez}. In the temperature range discussed here, the upper frequency limit for JNT is typically set by the amount of stray capacitance from the graphene, lead wires, and contact pads to the Si back gate. This is minimized by using short, narrow lead wires and small (50 $\mmu$m $\times$ 50 $\mmu$m) bonding pads resulting in an estimated 4 pF stray capacitance. 

The amount of Johnson noise emitted between any two terminals is proportional to the mean electronic temperature between them where each point in space is weighted by its local resistance. Therefore, to maximize the signal coming from the graphene, contact resistance should be kept at a minimum. To compensate for the narrow lead wires, we deposit a thicker layer (200 nm) of gold resulting in an estimated total contact resistance of $<80\;\Omega$.

Lastly, to effectively extract $\kappa_{\mathrm{e}}$ from the total electronic thermal conductance $G_{\mathrm{th}}$ we want to enhance the electron diffusion cooling pathway with respect to the electron-phonon cooling pathway (see below). This can be accomplished by keeping the length of the sample short as the total power coupled into the lattice scales as the area of the device while diffusion cooling scales as $1/R$. In addition, the device should be made wide to minimize the effects of disordered edges. We find these high aspect ratio samples ($\sim$ 3:1) are ideal for our measurements and serve the additional purpose of lowering the total sample resistance allowing us to impedance match over a wider bandwidth.

\begin{table}[t]
\begin{tabular}{| l | c | c | c |}\hline
&\  S1 &\ S2 &\ S3 \\\hline
length ($\mmu$m) &\ 3 &\ 3 &\ 4 \\
width ($\mmu$m) &\ 9 &\ 9 &\ 10.5 \\
mobility ($10^5\; \mathrm{cm}^2\cdot \mathrm{V}^{-1}\cdot \mathrm{s}^{-1}$) &\ $3$ &\ $2.5$ &\ $0.8$ \\
$n_{\mathrm{min}}$ ($10^9$ cm$^{-2}$) &\ 5 &\ 8 &\ 10 \\\hline
\end{tabular}
\caption{Basic properties of our three samples.}
\label{table1}
\end{table}

\section{Device Characterization}
\begin{figure}[t]
\includegraphics[width=3in]{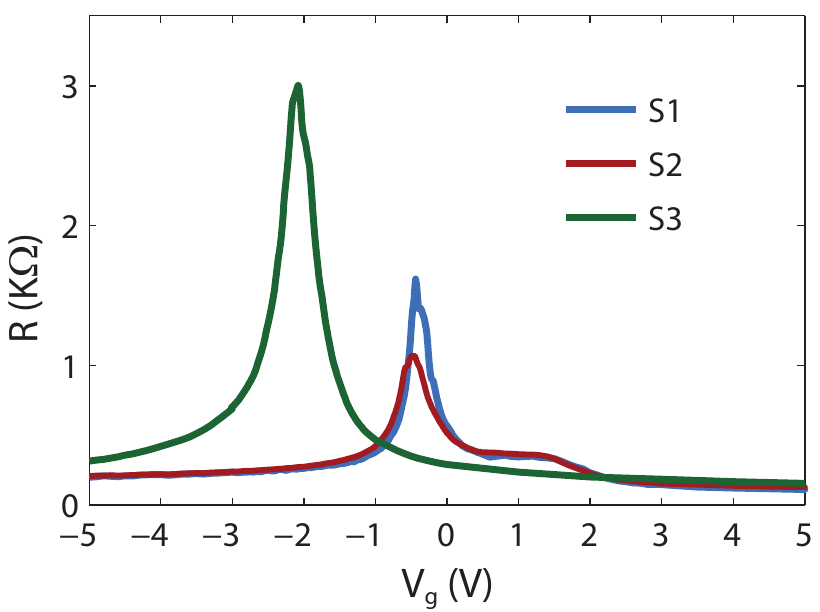}
\caption{2-terminal resistance $R$ vs. back gate voltage for the 3 samples used in this report.}
\label{figs1}
\end{figure}

In this study we measure three graphene devices encapsulated in hexagonal boron nitride (hBN), whose basic properties are detailed in Table \ref{table1}.    All devices are two-terminal with mobility estimated as \begin{equation}
\mu\approx \frac{L}{neRW},
\end{equation}
where $L$ and $W$ are the sample length and width respectively, $e$ is the electron charge, and $n$ is the charge carrier density.   The gate capacitance per unit area $C_{\mathrm{g}}\approx 0.11\; \mathrm{fF}/\mmu \mathrm{m}^2$ is estimated considering the 285 nm SiO$_2$ and $\sim 20$ nm hBN dielectrics.   From this we estimate the charge density \begin{equation}
n=\frac{C_{\mathrm{g}}(V_{\mathrm{g}}-V_{\mathrm{d}})}{e}
\end{equation} where $V_{\mathrm{d}}$ is the gate voltage corresponding to the charge neutrality point (CNP) estimated by the location of the maximum of the curve $R(V_{\mathrm{g}})$.   Fig. \ref{figs1} shows the resistance of all samples as a function of gate voltage.

\section{Johnson Noise Thermometry}
The full Johnson noise thermometry (JNT) setup used in this study is outlined in detail in \cite{Crossno:2015ez}.   Here, we only give a brief synopsis for completeness.

The electronic transport within a dissipative device can be determined by the high frequency noise power collected by a low noise amplifier as \begin{equation}
\left\langle V^2\right\rangle = k_{\mathrm{B}}T_{\mathrm{e}} \times \mathrm{Re}(Z)\Delta f \left[1-\left(\frac{Z-Z_0}{Z+Z_0}\right)^2\right]
\end{equation}where $Z$ is the complex impedance of the device under test, $Z_0$ is the impedance of the measure circuit (typically 50 $\Omega$) and $\Delta f$ is the bandwidth.   From this formula, we can see two critical components of JNT:  impedance matching over a wide bandwidth and low noise amplification.

\begin{figure}
\includegraphics[width=3in]{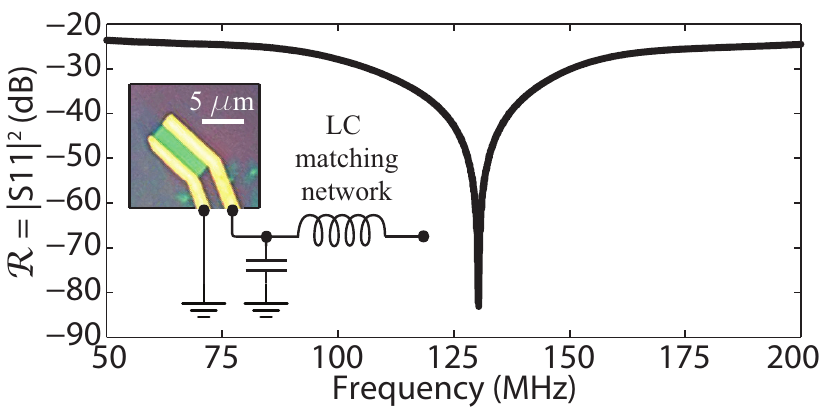}
\caption{Reflectance $\mathcal{R}=|S_{11}|^{2}$ for a graphene device impedance matched to 50 $\Omega$ near 125 MHz.}
\label{figs2}
\end{figure}

Graphene devices have a typical channel resistance on the order of $h/4e^2 \sim 6\; \mathrm{k}\Omega$ near the CNP.   To compensate for this, we use an inductor-capacitor (LC) tank circuit mounted directly on the sample package to impedance match the graphene to the 50 $\Omega$ measurement network.  Fig.  \ref{figs2} shows the reflectance coefficient $\mathcal{R}=|S_{11}|^{2}$ for a typical graphene device after impedance matching.   The bandwidth and measurement frequency of our JNT is set by the Q-factor and LC time of this matchingnetwork.

At 10 K, the power emitted by a resistor in a 1 Hz bandwidth is $\sim 10^{-22}$ W or $-190$ dBm.  To amplify this signal we use a SiGe low noise amplifier (Caltech CITLF3) with a room temperature noise figure of about 0.64 dB in our measurement bandwidth, corresponding to a noise temperature of about 46 K.   We operate the amplifier at room temperature, outside of the cryostat, to ensure it is unaffected by the 3--300 K temperature ramp used for thermal conduction measurements.  After amplification, a homodyne mixer and low pass filter define the measurement bandwidth and the power is found by an analog RF multiplier operating up to 2 GHz (Analog Devices ADL5931).  The result is a voltage proportional to the Johnson noise power which -- after calibration -- measures the electron temperature in the graphene device.

\begin{figure}
\includegraphics[width=3in]{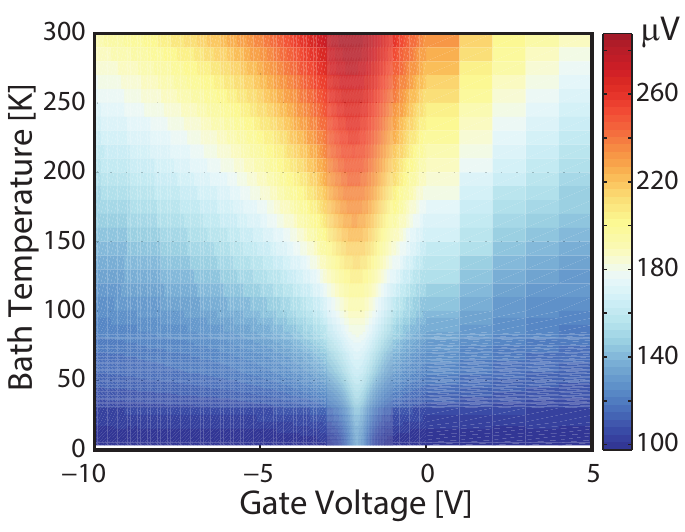}
\caption{Output voltage $V_{\mathrm{s}}$ from the JNT measuring a graphene device with no excitation current.  This is used to calibrate the JNT to a given sample.}
\label{figs3}
\end{figure}

Calibration of our JNT device must be done on every sample as each device has a unique $R(T,V_{\mathrm{g}})$ and therefore couples differently to the amplifier.  The graphene device being measured is placed on a cold finger in a cryostat with varying temperature $T_{\mathrm{bath}}$.  With no excitation current in the graphene, we collect the JNT signal $V_{\mathrm{s}}(T,V_{\mathrm{g}})$ for all temperatures and gate voltage needed in the study, as shown in Fig. \ref{figs3}.   The linear temperature slope at each point gives a gain factor $g(T,V_{\mathrm{g}}) = \partial V_{\mathrm{s}}/\partial T$.

\section{Measuring electronic thermal conductance}

\begin{figure}
\includegraphics[width=3in]{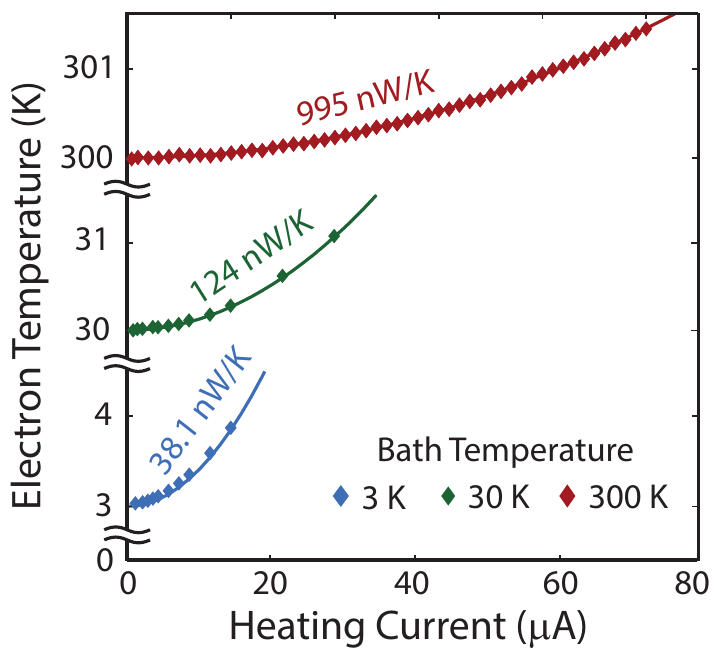}
\caption{The electronic temperature in an encapsulated graphene device as a function of heating current for three different bath temperatures.  $T_{\mathrm{e}}=T_{\mathrm{bath}}+I^2R/G_{\mathrm{th}}$.  The total thermal conductance between the electronic system and the bath is found through Fourier's Law (solid lines).}
\label{figs4}
\end{figure}

The procedure used to measure electronic thermal conductance is outlined in \cite{Crossno:2015ez, Fong:2012ut, prober}.  A small sinusoidal current $I(\omega)$ is run through the graphene sample, causing a Joule heating power $P(2\omega)$ to be injected directly into the electronic system.  This causes a small temperature difference $\Delta T(2\omega) $ between the electronic temperature and the bath, described by Fourier's law: \begin{equation}
P = G_{\mathrm{th}} \Delta T.   \label{eqp1}
\end{equation}
Here $G_{\mathrm{th}}$ is the total thermal conductance between the electronic system and the bath.  The component of Johnson noise at frequency $2\omega$  is measured by a lock-in amplifier and then converted to a temperature difference $\Delta T$ using the gain $g(T,V_{\mathrm{g}})$ described in the previous section.  Fig. \ref{figs4} shows $T_{\mathrm{e}}$ as a function of heating current $I$ for a graphene device at three different bath temperatures:  3, 30 and 300 K.

\section{Thermal model of graphene electrons}\label{s6sec}
\begin{figure}
\includegraphics[width=1in]{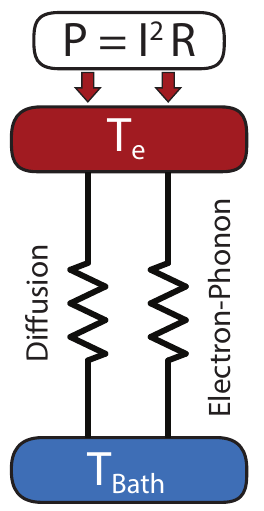}
\caption{Simplified thermal diagram of the electronic cooling pathways in graphene relevant for our experimental conditions.  A current induces a heating power into the electronic system which conducts to the bath via two parallel pathways:  diffusion and coupling to phonons.}
\label{figs5}
\end{figure}

In the regime presented here, $G_{\mathrm{th}}$ is dominated by two electronic cooling pathways.  Hot electrons can diffuse directly out to the contacts ($G_{\mathrm{diff}}$), or they can couple to phonons ($G_{\mathrm{el-ph}}$):  \begin{equation}
G \approx G_{\mathrm{diff}} + G_{\mathrm{el-ph}}.
\end{equation}
In a typical metal, electron diffusion is described by the WF law which is linear in $T_{\mathrm{e}}$.  The electron-phonon cooling pathway has two components:  first, the electrons must transfer heat to the lattice via electron-phonon coupling, and then the lattice must conduct the heat to the bath.   Consistent with previous experimental studies on graphene \cite{Crossno:2015ez, Fong:2013hl, Yigen:2014vw, hakonen}, we find $G_{\mathrm{el-ph}}$ is bottlenecked by the weak electron-phonon coupling and hence the lattice is well thermalized to the bath.  Fig. \ref{figs5} shows the simplified thermal diagram of the electronic cooling pathways in graphene, relevant to our experiment.

\begin{figure}
\includegraphics[width=3in]{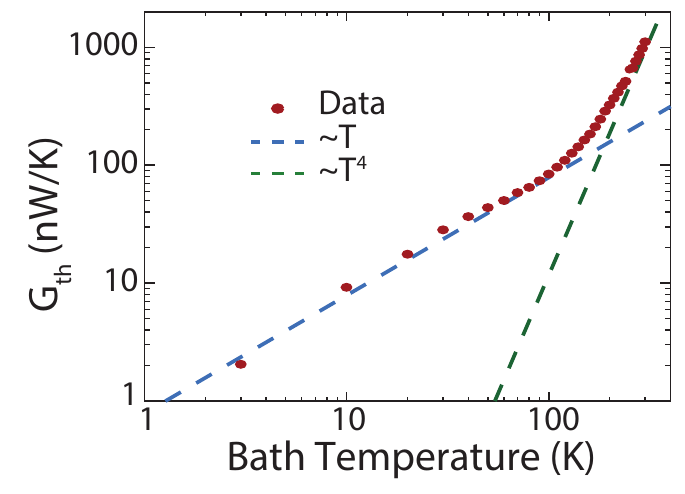}
\caption{The total thermal conductance of a graphene device in the high density regime, illustrating how $G_{\mathrm{th}}$ is dominated by either $G_{\mathrm{diff}}\sim T$, as predicted by the WF law, or $G_{\mathrm{el-ph}}\sim T^4$.}
\label{figs6}
\end{figure}

At low temperature, $G_{\mathrm{th}}$ is dominated by $G_{\mathrm{diff}}$, while at high temperature it is dominated by $G_{\mathrm{el-ph}}$.  Fig. \ref{figs6} shows an illustration of this effect in one of our devices.

\section{Measuring Thermal Conductivity}
In the diffusion-limited regime, we extract the electronic thermal conductivity $\kappa_{\mathrm{e}}$ as follows.    This is detailed in \cite{Fong:2012ut} and we review the basic calculation for clarity.   The total power dissipated, $P$, is given by \begin{equation}
P = \frac{J^2}{\sigma} LW = \mathcal{P}LW  \label{eqp2}
\end{equation}
where $L$ is the length of the sample in the direction of current flow,  $W$ is the width in the perpendicular direction and $\mathcal{P}$ is the local power dissipated.  Because this calculation is done in linear response, and the external heat baths on either side of the sample are at the same temperature,  the contributions to power dissipated $\sim \kappa (\Delta T)^2$ do not enter so long as $\Delta T \sim J^2$ is  small.   This is an appropriate assumption in the regime of linear response, where $J$ is treated as a perturbatively small parameter.    Fig. \ref{figs4} shows our experiment is in this regime.

Let us now determine the change in the temperature profile.   For simplicity we assume that the graphene sample is homogeneous, that the approximately uniform electrical current is given by \begin{equation}
J = -\sigma \frac{\mathrm{d}V}{\mathrm{d}x} - \alpha \frac{\mathrm{d}T}{\mathrm{d}x},
\end{equation}
and that the heat current is given by \begin{equation}
Q = -\alpha T\frac{\mathrm{d}V}{\mathrm{d}x} - \bar\kappa_{\mathrm{e}} \frac{\mathrm{d}T}{\mathrm{d}x},
\end{equation}
where \begin{equation}
\bar\kappa_{\mathrm{e}} \equiv \kappa_{\mathrm{e}} + \frac{T\alpha^2}{\sigma} = \kappa_{\mathrm{e}}(1+ZT).
\end{equation}
In the latter equation,  $ZT$ is the thermoelectric coefficient of merit.    As $\alpha \approx 0$ at the CNP,  we expect $ZT\approx 0$,  and that $\bar\kappa_{\mathrm{e}}\approx \kappa_{\mathrm{e}}$. 

 $\mathrm{d}T/\mathrm{d}x$ is the temperature gradient in the sample,  and $-\mathrm{d}V/\mathrm{d}x$ is the electric field in the sample.  $\alpha/\sigma$ is the Seebeck coefficient.  We assume that the response of graphene is dominated only by the changes in voltage $V$ and temperature $T$ to a uniform current density $J$, which is applied externally.  We also assume that deviations from constant $V$ and $T$ are small, so that the linear response theory is valid.  Joule heating leads to the following equations: \begin{subequations}\begin{align}
0 &= \frac{\mathrm{d}J}{\mathrm{d}x}, \\
\mathcal{P} = \frac{J^2}{\sigma} &=  \frac{\mathrm{d}Q}{\mathrm{d}x},
\end{align}\end{subequations}which can be combined to obtain \begin{equation}
\mathcal{P} = -\kappa_{\mathrm{e}} \frac{\mathrm{d}^2 T}{\mathrm{d}x^2},
\end{equation}
assuming that $\kappa_{\mathrm{e}}$ is approximately homogeneous throughout the sample.

 \begin{figure}
\includegraphics[width=3.5in]{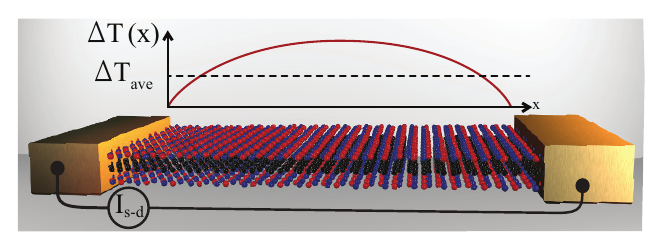}
\caption{Cartoon illustrating the non-uniform temperature profile within the graphene-hBN stack during Joule heating in the diffusion-limited regime.}
\label{figs7}
\end{figure}

The contacts in our experiment are held at the same temperature $T$.   Thus, writing \begin{equation}
T(x) = T_{\mathrm{e}} + \Delta T(x),
\end{equation}we find that \begin{equation}
\Delta T(x) = \frac{\mathcal{P}}{2\kappa_{\mathrm{e}}}x(L-x).  
\end{equation}
The average temperature change in the sample, which is directly measured through JNT, is \begin{equation}
\langle \Delta T\rangle = \int\limits_0^L \frac{\mathrm{d}x}{L}\; \Delta T(x)  = \frac{\mathcal{P}L^2}{12\kappa_{\mathrm{e}}}.  \label{eqp3}
\end{equation}
This non-uniform temperature profile is illustrated in Fig. \ref{figs7}.   Combining Eqs. (\ref{eqp1}), (\ref{eqp2}) and (\ref{eqp3}) we obtain \begin{equation}
G_{\mathrm{th}} = \frac{12L}{W} \kappa_{\mathrm{e}}.
\end{equation}

As we have pointed out in the main text,  our samples are not perfectly homogeneous, but have local fluctuations in the charge density.   Nevertheless, we do recover the WF law in the FL regime, suggesting that our measurement of $G_{\mathrm{th}}$ -- and thus $\kappa_{\mathrm{e}}$ -- using JNT, along with the above formalism, is valid.

%\pagebreak
%\bibliographystyle{unsrt}
\bibliography{HydroGrapheneV14}
\end{document}